
\input phyzzx
\baselineskip 13pt

\date{LPTHE 93/41 October 1993}
\titlepage

\title{SPACE-TIME EVENTS AND RELATIVISTIC PARTICLE LOCALIZATION}
\author{J. Mourad\foot{e-mail:MOURAD@QCD.UPS.CIRCE.FR}}
\address{{ Laboratoire de Physique Th\'eorique et Hautes
Energies}\foot{Laboratoire associ\'e au CNRS.},
 Universit\'e de Paris-Sud, B\^at. 211, 91405 Orsay
Cedex, France}
\abstract{ A relation expressing the covariant transformation properties of a
relativistic position operator is derived.
This relation differs from the one existing
in the literature expressing manifest
covariance by some factor ordering. The relation is derived in
order for the localization of a particle to represent a space-time event.
It is shown
that there exists a conflict between this relation and the hermiticity
of a positive energy position operator.}


\endpage
\chapter{Introduction}

In classical physics one has a realization of the concept
of a space-time event [1]. Namely, the fact that a particle is localized in
$\bf x$ at time $t$ represents the space-time event $x$. The aim
of this article is to investigate whether one can have in quantum mechanics,
at least in principle,
the same situation. Can the fact that a particle is localized
in space at a given time represent a space-time event? A space-time
event is characterized, in a flat space, by its transformation
properties under the Poincar\'e group,
$$x^{\mu} \rightarrow \Lambda^{\mu}_{\nu}x^{\nu}+a^{\mu}.$$

On the other hand, a particle, in quantum theory, is defined as
a realization of an irreducible unitary representation of
the Poincar\'e group [2]. It seems difficult to include interactions
in a relativistic invariant manner
 for these irreducible representations. So the particle is described, at
 the first quantized level,
 by
 a reducible representation having both signs of the energy.
 At the second quantized level, positive energy irreducible representations
 characterize the one particle sector of the Hilbert space
 of the theory.

 The requirement of relativistic
invariance says nothing about the localization properties of
the particle. We have a Hilbert space carrying a representation
of the Poincar\'e group and all the operators that act on this
space. How can we construct observables in this space?

In non-relativistic
quantum theory we have a more or less well defined procedure: the
principle of equivalence. There one begins with the fundamental
classical observables, the position and the momentum and turns them into
operators satisfying the Heisenberg commutation relations but having
the same physical meaning of the classical observables. A
way to solve the problem in a relativistic theory is to use the
geometrical origin of the Poincar\'e transformations and
to impose on the observables certain constraints arising
from their preimposed transformation properties and hope that these do
determine the observables completely. This is the way that
relativistic position operators have been determined. For
instance in reference [3] it was shown that the requirement that the
position operator transforms in
a given way under translations, rotations, space and time inversions
determines it completely if one searches for a self-adjoint operator
with commuting components.

Pryce [4] and later on Currie, Jordan and Sudarshan [5],
 in the same spirit, gave relations
that characterize covariant position operators. However they
used a semi-classical reasoning. They found the classical relations
and then converted the Poisson Brackets into commutators with a given ordering
prescription, namely they used the symmetric ordering of
operators.

The aim of this article is to reexamine this relation in order to see
whether this ordering prescription is justified. In order to do that
we rely on our above definition of covariance. Namely a position
operator is defined to be manifestly covariant if its
eigenstates transform under a pure Lorentz transformation as
does a space-time event.

In section $2$ classical manifest covariance is briefly reviewed.
The transformation of the Newton-Wigner eigenstates under a boost
is the subject of section $3$.
In section $4$
 the relation expressing the covariance of the quantum localization is
derived. The solution to this relation, for the scalar particle, is
discussed in section $5$.
Finally we conclude with some general and speculative remarks.

\chapter{Classical manifest covariance}

 In order to have relativistic invariance it is
necessary and sufficient to have a collection of generators
$J_{i},K_{i}$ representing the Lorentz Lie algebra on phase space by
means of the Poisson Brackets.
Relativistic invariance does not dictate the functionnal form of
the transformation of the position
variables. We can add the requirement of covariant
transformation by giving the form of the transformed coordinates. We will
do that in the following.

Suppose the particle is described by the position $q_{i}(t)$ in a
given inertial frame $\cal R$.

In a frame related to the first by an infinitesimal Lorentz
transformation the position is described by
$$ q_{j}'(t)=q_{j}(t)- \epsilon^{i}\{K_{i},q_{j}(t)\} \eqno(2.1)$$

Note that time is not transformed in this equation.

In order to have covariant transformation properties of the position we
must have

$$ \eqalign{
q_{j}'(t')=q_{j}(t)-\epsilon_{j}t,\cr
t'=t-\epsilon^{i}q_{i}(t).\cr}\eqno (2.2)$$

These relations simply state that the fact that the particle is localized
in a given position at a given time reprsents a space-time event.
{}From  relations $(2.2)$ one can deduce how the positions in the two
frames are related at the same time, to first order in $\epsilon$,

$$q_{j}'(t)=q_{j}'(t')-(t-t')\{H,q_{j}'(t')\}=q_{j}(t)-\epsilon_{j}t -
\epsilon^{i}q_{i}(t)\{H,q_{j}(t)\}.\eqno (2.3)$$
$H$ is the Hamiltonian of the system.
Comparing the two equations $(2.1)$ and $(2.3)$
 we find the relation,

$$\{K_{j},q_{i}\}=t\delta_{ij}+q_{j}\{H,q_{i}\}. \eqno (2.4) $$
This is the relation expressing the covariance of the classical position
variable [4,5]. Define $N_{j}$ as usual by
$$N_{j}=K_{j}+tp_{j},\eqno (2.5)$$
where $p_{j}$ is the translation generator satisfying $\{q_{j},p_{i}\}=
\delta_{ij}$. Our final relation becomes,
$$\{N_{j},q_{i}\}=q_{j}\{H,q_{i}\}.\eqno (2.6)$$
A solution to this equation, for the spinless
particle, satisfying as well the correct
transformation properties under the other Poincar\'e generators is
$$q_{i}={N_{i} \over {H}}.\eqno (2.7)$$

\chapter {Transformation of the Newton-Wigner eigenstates}

The Newton-Wigner position operator was introduced as the unique
positive energy operator satisfying the axioms of localizability
of Newton and Wigner [3,6]. These axioms require the commutativity of the
components, the hermiticity, and the correct transformation properties under
the translations, rotations, parity and time reversal. The
Newton-Wigner operator does not exist for massless spinning particles.
It plays an important role in the study of the non-relativistic
limit [7] and the semi-classical approximation [8].
The transformation under boosts of the Newton-Wigner operator
were not postulated and is the subject
of this section.

Consider a free scalar particle, a basis of its Hilbert space
is given by the momentum eigenstates so that
a general state may be written as
$$ |\psi>=\int d\tilde {\bf p} \  \psi({\bf p})\ |{\bf p}>.\eqno (3.1)$$

Here the measure is the invariant one,

$$d\tilde {\bf p}={d^{3}p \over {(2 \pi)^{3 \over {2}} \omega({\bf p})}},
\ \omega({\bf p})=\sqrt{{\bf p}^{2}+m^{2}}.\eqno (3.2)$$

The scalar product is such that,
$$<{\bf p}|{\bf p}'>=(2 \pi)^{3 \over {2}} \omega({\bf
p})\delta^{3}({\bf p-p'}).\eqno (3.2)$$
This reproduces the scalar product of Bargmann and Wigner [9],
$$<\phi|\psi>=\int d\tilde {\bf p} \ \phi^{*}({\bf p}) \psi({\bf p}).
\eqno (3.3)$$

The eigenstates of the Newton-Wigner position operator are given
by,
$$ \psi_{\bf q_{0}}({\bf p})=\sqrt{\omega({\bf p})}e^{-i{\bf p.q_{0}}},
\eqno (3.3)$$
this state represents a particle localized in $\bf q_{0}$ in a frame
$\cal R$ at time $t$. We are interested in the state of
the particle viewed from a frame $\cal R'$, related to $\cal R$
by a boost with velocity $\bf \beta$. Let $t',{\bf q_{0}}'$  be the Lorentz
transform of $t,{\bf q_{0}}$.
The state at time $t'$
is given, in configuration space, by,
$$ <{\bf q}|\psi>'_{t'}=
\psi'({\bf q})={\sqrt{\gamma} \over {(2\pi)^{3}}} \int d{\bf p}
\sqrt{\left(1-{{\bf \beta.p} \over {\omega}}\right)}e^{i{\bf
p.(q-q_{0}')}}.
\eqno (3.4)$$
Here, $\gamma$ is defined by,
$$ \gamma=\sqrt{1-{\bf \beta}^{2}},\eqno (3.5)$$
and $|{\bf q}>$ is an eigenstate of the Newton-Wigner operator.

The function  $(3.4)$ cannot vanish outside a bounded domain because
it is the Fourier transform of a non analytic function. This is due
to the presence of square roots in the integrand.
 The
state in the frame $\cal R'$ is not localised at time $t'$. One may
easily verify that it is not
localised at any time. This proves that the Newton-Wigner operator
is not manifestly covariant. The integral may
be exactly evaluated in the massless
one-dimensional case. The result is,
$$\psi'(q)={\sqrt{\gamma} \over {2}}\left(\sqrt{1+\beta}+\sqrt{1-\beta}
\right)\delta(q-q_{0}')+{i\sqrt{\gamma} \over {2\pi}}
\left(\sqrt{1-\beta}-\sqrt{1+\beta}\right){PP \over {q-q_{0}'}},\eqno (3.6)$$
where $PP$ denotes the principal part.

We conclude that the localisation, in the Newton-Wigner sense,
of a particle at a given time does not represent a space-time
event. This fact has been noted several times in the literature [3,10,11],
but to the knowledge of the author no explicit calculation has been
given.

\chapter{Quantum manifest covariance}

In this section we are interested in the relation expressing the
covariance property of the quantum relativistic position operator.
It is known that many relativistic position
operators considered in the literature [4,12,13]
have non-commuting components. So one may argue that localization may
be meaningless in a relativistic theory, why then search
for a covariant localization? The point is that although
the position operator may have non-commuting
components, one may find, as was recently been proved for
massless particles [12], states localized within an arbitrary
small region in space. The question is now whether this localization
is covariant or not. The effect of having non-commuting
components will be neglected in the following, but
this may easily be cured by considering states localized in an
arbitrary small region
instead of exactly localised states. Anyway, we may restrict
ourselves to scalar particles and verify at the end that a
solution having commuting components exists.

Suppose the particle is described by a state vector localized in $\bf x$
at time $t$ in an inertial frame $\cal R$,
$$ |\psi>_{t}=|\bf x> \otimes |\alpha>.\eqno (4.1)$$
Where $\alpha$ represents internal degrees of freedom (spin,$\dots$).
The observer in a frame $\cal R'$ related to $\cal R$ by an
 infinitesimal boost
would describe the particle, at the instant $t$, by the
state vector
$$|\psi>_{t}'= |\psi>_{t}-i\epsilon^{i} \hat K_{i}|\psi>_{t}. \eqno(4.2)$$

Note that, as in the classical case, time is unchanged.

We say that we have manifest covariance if the observer in $\cal R'$ sees the
particle localized in $\bf x'$ at time $t'$, the latter being related
to $\bf x$ and $t$ by a Lorentz transformation. In other
words manifest covariance is the requirement that the localized particle
at a given time represents a space-time event. This is realised
if we have
$$|\psi>_{t'}'=|\bf x'>\otimes|\alpha'>.\eqno(4.3)$$
Since we are interested only in the manifestly covariant localization
we do not impose any requirement on the internal variables
$\alpha'$. Our aim is now to find
the relation satisfied by the position operator in order
for this to be true. We will have to translate in time the
state vector $|\psi>_{t}'$ to get it at time $t'$ and then impose
the relation $(4.3)$.
The particle is described at time $t'$ in the frame $\cal R'$ by
the wavefunction given by
$$|\psi>_{t'}'=|\psi>_{t}'-i(t'-t) \hat H|\psi>_{t}'.\eqno (4.4)$$
Here $\hat H$ is the Hamiltonian of the particle. Using
the relation (4.2) and the value for $t'-t$ given in $(2.2)$
 we get, to first order in $\epsilon$,

$$|\psi>_{t'}'=|\psi>_{t}-i\epsilon^{j}\hat K_{j}|\psi>_{t}+
i\epsilon^{j}\hat H \hat x_{j}|\psi>_{t}.\eqno(4.5)$$
We have used the fact that the particle is localised in $\cal R$
in order to get the last term on the right hand side of the above relation.
We now  impose  relation $(4.3)$ by requiring
$$\hat x_{i}|\psi>_{t'}'=x_{i}'|\psi>_{t'}'=(x_{i}-\epsilon_{i}t)
|\psi>_{t'}'.\eqno (4.6)$$
Substituting for $|\psi>_{t'}'$  in the above relation
its value given by $(4.5)$ we get
the following relation,
$$\left( \hat x_{i}\hat K_{j}-\hat x_{i}\hat H\hat x_{j} \right)|\psi>_{t}=
\left(-i \delta_{ij}t+x_{i}\hat K_{j}-x_{i}\hat H \hat x_{j}
\right)|\psi>_{t}.
\eqno (4.7)$$
We now  replace $x_{i}|\psi>_{t}$ by $\hat x_{i}|\psi>_{t}$ and we
use the definition  $\hat N_{i}=\hat K_{i}+t\hat p_{i}$ to get the
required relation,

$$ [\hat x_{i},\hat N_{j}]=[\hat x_{i},\hat H \hat x_{j}].\eqno(4.8)$$
This is the main result of this article.

It is important to note that relation $(4.8)$ is realized for
a self-adjoint position operator only if the latter commutes with
$[\hat H,\hat x_{i}]$. This is realized for the Dirac
position operator. A solution to equation $(4.8)$, for a
scalar particle, satisfying
as well the correct transformation properties under the
other Poincar\'e generators is given by,
$$ \hat X_{i}={1 \over {\hat H}}\hat N_{i}.\eqno(4.9)$$

This operator is not self-adjoint since $\hat H$ and
$\hat N_{i}$ do not commute.
 The physical relevance of having non
self-adjoint position operators will be discussed in the next section.
An important property of the above operator is that it does not
couple positive and negative energy states because it is
constructed out of
the generators of the Poincar\'e group. Another property of this
operator is that it has commuting
components.

\chapter{Discussion}

Relation $(4.8)$ is not the one given by Pryce [4] and used later
to express the manifest covariance of position operators [5,10].
The implicit derivation of Pryce, which later was made more explicit by Currie,
Jordan and Sudarshan [5] used the principle of equivalence. The method
consisted in considering the classical relation and
converting the Poisson Bracket into a commutator
and finally ordering the products symmetrically. When one does
this one gets the following relation
$$ [\hat x_{i},\hat N_{j}]={1 \over {2}}\left(\hat x_{j}[\hat x_{i},\hat H]
+[\hat x_{i},\hat H] \hat x_{j}\right).\eqno (5.1)$$

Observe that this ordering prescription is somewhat arbitrary
and does not reproduce our result $(4.8)$. It coincides with it
when $\hat x_{i}$ commutes with $[\hat x_{j},\hat H]$. This is the case
for the Dirac position operator.

We now discuss further the solution to relation $(4.8)$ for the
spin 0 case.

 Consider a free scalar particle satisfying the
Klein-Gordon equation:
$$ \left( p^{2}-m^{2} \right) \phi (p)=0.\eqno (5.2)$$
We use the Hilbert space scalar product $(3.3)$.

With this scalar product the Newton-Wigner position operator is
given by [3],
$$\hat q_{i}=i{\partial \over {\partial p_{i}}} -i{p_{i} \over {2\omega(p)^{
2}}}.\eqno (5.3)$$
Note that this operator is self-adjoint with respect
to  the scalar product $(3.3)$.
The covariant position operator $(4.9)$ is given by,
$$\hat X_{i}=i{\partial \over {\partial p_{i}}}.\eqno (5.4)$$
It is not self-adjoint, because it differs from the self-adjoint
Newton-Wigner operator by a complex operator. The Newton-Wigner
operator is the self-adjoint part of $\hat X_{i}$,
$$ \hat q_{i}={1 \over {2}}\left( \hat X_{i}+\hat X_{i}^{\dagger}
\right). \eqno(5.5) $$
One can construct the eigenstates of the covariant position operator
in the momentum representation and find,
$$|{\bf x}>=\int d\tilde {\bf p} e^{-i{\bf x. p}}|{\bf p}>.\eqno (5.6)$$
This state is a solution to the equation
$$\hat X_{i}|{\bf x}>=x_{i}|{\bf x}>.\eqno (5.7)$$
Note that this state is obtained, at the second quantized
level, by applying the field operator $\hat \phi^{\dagger}({\bf x})$
to the vaccum
state,
$$ |{\bf x}>=\hat \phi^{\dagger}({\bf x})|vac>. \eqno(5.8)$$
The scalar product of two such eigenstates is given by,
$$<{\bf x '}|{\bf x}>=\int d\tilde{\bf p} \ e^{-i{\bf p}.({\bf x- x'})}.
\eqno (5.9)$$
The scalar product of two eigenstates with two different eigenvalues is
not zero. This is a manifestation of the non self-adjointness of the
covariant position operator. The usual interpretation of the scalar
product as a probability amplitude seems to exclude the use of non hermitian
operators as observables. Another manifestation of the
non-self-adjointness of the
covariant position operator
is that we no longer have the usual closure relation in
terms of the eigenstates
of the position operators and their duals. The closure relation may be
deduced from the relation
$$\hat 1=\int d\tilde {\bf p} \ |{\bf p}><{\bf p}|.\eqno (5.10)$$
After inverting relation $(5.4)$ to get the momentum
eigenstates in terms of the position eigenstates, we get the closure
relation,
$$\hat 1=\int \ d{\bf x}d{\bf x'}\ |{\bf x}>\tilde \delta ({\bf x}-{\bf x'})
<{\bf x'}|,\eqno (5.11)$$
where the function $\tilde \delta ({\bf x}-{\bf x'})$ is defined by,
$$\tilde \delta ({\bf x}-{\bf x'})=\int {d\tilde {\bf p} \over
{(2 \pi)^{3}}} \ \omega(p)^{2} e^{i{\bf p.(x'-x)}}.\eqno (5.12)$$

Using this closure relation  we can get the scalar product in
the covariant position representation,
$$<\psi|\psi>=\int d{\bf x}d{\bf x'}\ \psi({\bf x'})^{*} \tilde
\delta ({\bf x}-{\bf x'}) \psi({\bf x}).\eqno (5.13)$$
We can see from the above relation that the
interpretation of $|\psi({\bf x})|^2 \equiv <\psi|x><x|\psi>$
as a probability density
is not possible
since the probabilities do not add to one.

One is thus forced to conclude that the eigenstates of this covariant
position operator cannot represent physical localized states.

\chapter{Conclusion}

We have shown the impossibility of constructing localized
states that transform under Lorentz transformations as
do space-time events. We conclude that in a quantum theory
space-time events do not have a physical realization.
The motivation for introducing  space-time was a classical one [1].
In a quantum theory, space-time seems not to have an exact physical
reality. Its role is reduced to a parametrization of
fields, permitting some important concepts such as
gauge invariance and locality. One is tempted to abandon the
manifold structure of space-time
for other constructions that reduce to it at large length scales [14].

\beginsection {\bf Acknowledgements}

I am indebted to  H. Sazdjian for many helpful conversations.

\beginsection{\bf References}

\item{[1]} see e.g. H. Minkowski, Address delivered at the 80th Assembly
of German Natural Scientists and Physicians, Cologne, 1908.

\item {[2]} E.P. Wigner, Ann. Math. 40 (1939) 149.

\item {[3]} T.D. Newton and E.P. Wigner, Rev. Mod. Phys. 21 (1949) 400.

\item {[4]} M.H.L. Pryce, Proc. Roy. Soc. (London) A195 (1948) 62.

\item {[5]} D.G. Currie, T.F. Jordan, E.C.G. Sudarshan,
Rev. Mod. Phys. 35 (1963) 350.

\item {[6]} A.S. Wightman, Rev. Mod. Phys. 34 (1962) 845.

\item {[7]} L.L Foldy and S.A. Wouthuysen, Phys. Rev. 78 (1950) 29.

\item {[8]} J. Mourad, Phys. Lett. A 179 (1993) 231.

\item {[9]} V. Bargmann and E.P. Wigner, Proc. Nat. acad. of sci. 34 (1946)
211.

\item {[10]} T.F. Jordan and N. Mukunda, Phys. Rev. 132 (1963) 1842.

\item {[11]} G.N. Fleming, Phys. Rev. 137 (1965) B188; 139 (1965) B963.

\item {[12]} J. Mourad, Optimal photon localization,
to be published in Phys. Lett. A.

\item {[13]} H. Bacry, Annls. Inst. H. Poincar\'e 49 (1988) 245.

\item {[14]} see e.g. H.S. Snyder, Phys. Rev. 71 (1947) 38.

\end